\documentclass[twocolumn,aps]{revtex4}

\usepackage{graphicx}% Include figure files
\usepackage{dcolumn}% Align table columns on decimal point
\usepackage{bm}% bold math

\begin{document}
%\preprint{APS/123-QED}

\title{Strain sensing with sub--micron sized Al--AlO$_{\mathrm{x}}$--Al tunnel junctions}
% Force line breaks with \\

\author{P.~J. Koppinen}
%\email{panu.koppinen@jyu.fi}
\author{J.~T. Lievonen}
\author{M. Ahlskog}
\author{I.~J. Maasilta}
\affiliation{Nanoscience Center, Department of Physics, University of Jyv\"askyl\"a, P. O. Box 35, 
FI--40014 University of Jyvä\"askyl\"a, Finland.}

\date{\today}% It is always \today, today,
             %  but any date may be explicitly specified

\begin{abstract}
We demonstrate a local strain sensing method for nanostructures based on metallic Al tunnel junctions with AlO$_{\mathrm{x}}$ barriers. The junctions were fabricated on top of a thin silicon nitride membrane, which was actuated with an AFM tip attached to a stiff cantilever. A large relative change in the tunneling resistance in response to the applied strain (gauge factor) was observed, up to a value 37. This facilitates local static strain variation measurements down to $\sim 10^{-7}$. %In addition, we can show that strain can be measured locally with tunnel junctions and that tunneling resistance has effect to the response.
\end{abstract}

%\pacs{}% PACS, the Physics and Astronomy
                             % Classification Scheme.
%\keywords{Suggested keywords}%Use showkeys class option if keyword
                              %display desired
\maketitle
%\section{Introduction}
Tunnel junctions with aluminium oxide (AlO$_{\mathrm{x}}$) barriers are widely used in nanoelectronics applications, such as single electron transistors (SET)\cite{NATO}, SQUIDs\cite{vanDuzer}, radiation detectors\cite{enss}, superconducting quantum bits \cite{martinisreview}, thermometers and coolers \cite{jukkareview, SuspCooler}. However, their use in local mechanical strain and displacement sensing has been very limited, one rare example being  %anoscale strain sensitive tunnel junctions have been investigated, for example, in a case where the resistance of  
measurements of the %tunneling contact between a metal electrode and a carbon nanotube is modulated by  
mechanical vibrations of a nanotube\cite{Make}. In this letter, we show that conventional sub--micron sized aluminum tunnel junctions with AlO$_{\textrm{x}}$ barriers can be used for very sensitive strain and displacement detection in nanostructures. If used as a displacement detector, our scheme is much simpler than the standard scanning tunneling microscopy techniques, as the sensor is fabricated on the mechanical structure without a need to position and control an external electrode with a vacuum gap barrier.   %To our knowledge, the strain sensing has not been demonstrated with conventional sub--micron sized Al--AlO$_{\mathrm{x}}$--Al. 
%However, experiments with magnetic tunnel junctions has been made \ref{Siemens, Kiina}. 

In typical applications, microscale displacement detection is based either on capacitive or optical sensing\cite{ClelandNEMS}. Both of these methods have limitations in the sub-micron scale: the capacitance of a device can be dominated by parasitic capacitances of the measurement setup and optical sensing is mainly limited by diffraction. In addition, there are several other methods for strain and displacement sensing based on the phenomenon of piezoresistivity\cite{Piezo1,Piezo2,Piezo3}. Typically semiconductors or thin metal films are used as piezoresistive transducers. In addition, a few groups have reported applicability of magnetic tunnel junctions for strain sensing \cite{Siemens,Kiina}. However, all methods mentioned above have some limitations regarding to possible applications: semiconducting films are not very suitable for high frequency applications due to their high resistivity \cite{Roukes}, %which makes the impedance matching difficult,
 metal films have a relatively low response to the applied strain\cite{Roukes}, and magnetic tunnel junctions require external magnetic fields. Also, up to date the magnetic tunnel junction sensors have been relatively large (20 x 20 $\mu$m$^2$)\cite{Siemens}. 

Here, we stress that lithographically fabricated  tunnel junction devices have several advantages in strain detection, such as their small size (dimensions are easily made to be $<$100 nm), existing high--frequency read--out schemes\cite{RFSET,RFSINIS}, ease of integration into a mechanical system and a good response to the applied strain. In addition, the measurement scheme is very simple, as only a measurement of the change in the tunneling resistance is needed.

Our experimental setup consisted of a thin (30 nm or 65 nm) silicon nitride (SiN) membrane (dimensions 250 $\mu$m $\times$ 250 $\mu$m $\times$ 65 nm or 120 $\mu$m $\times$ 150 $\mu$m $\times$ 30 nm), onto which tunnel junctions of lateral dimensions 200 nm $\times$ 300 nm were fabricated with conventional electron beam lithography and multi-angle vacuum evaporation techniques\cite{Phonons2007}. The membrane was actuated by an atomic force microscope (AFM) tip and the tunneling resistance was measured simultaneously with standard lock--in techniques. In addition to the tunneling resistance, the photodetector signal and the z--piezo movement, i.e. the vertical displacement of the AFM cantilever were recorded. All measurements presented in this paper were performed at ambient conditions.

\begin{figure}
\includegraphics[width=6cm]{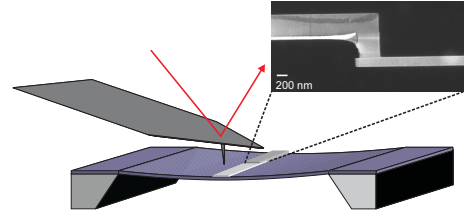}
\caption{\label{fig:schema} (Color online) Schematic view of the experimental setup. Zoom--in: SEM image of a tunnel junction. Note that the plane of the junction is parallel to the membrane.}
\end{figure}

An AFM tip works as a local actuator for the membrane, and therefore provides the possibility to measure the in-plane spatial dependence of the strain field. In addition, the AFM method provides a high displacement accuracy, particularly in the vertical direction. Strain was applied to the membrane by pressing it slowly down to a predetermined depth in 100 seconds (usually $\sim 3 \mu$m), keeping the maximum  displacement for 60 seconds, and then releasing it back to its original shape in another 100 seconds. Similar actuation scheme has been recently used to study the elastic properties of thin graphene sheets\cite{Sciencegraphene,VanDerZantgraphene}.  Stiff cantilevers \cite{note} were used in order to minimize the bending of the cantilever during actuation, and thus exerting most of the force to the membrane. From the measured force vs. displacement (FZ) curves the actual vertical displacement of the membrane was calculated by substracting the bending of the cantilever from the z--piezo movement, as described in Refs. \cite{Sciencegraphene,VanDerZantgraphene}. A schematic of the experimental setup and an SEM image of a typical tunnel junction are shown in Fig. 1. 
\begin{figure}
\includegraphics[width=8.7cm]{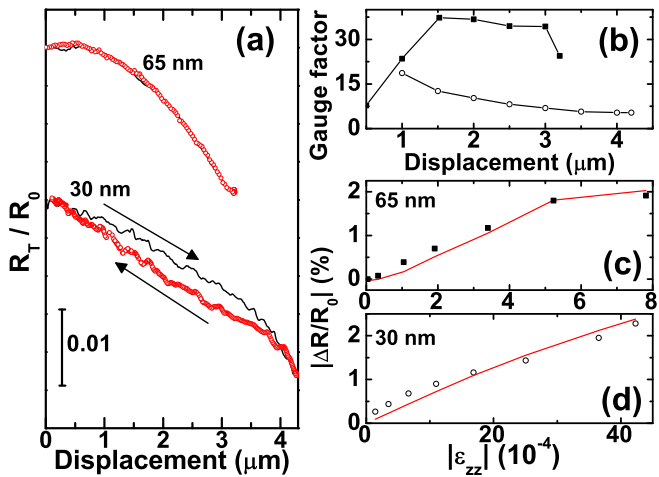}
\caption{\label{fig:response}(Color online) (a) Tunneling resistance $R_{T}$ normalized with unstrained (zero displacement) resistance $R_{0}$ as a function of vertical displacement. Press and release phases are shown with black solid line and red open circles, respectively. Top curve is data for a 65 nm thick membrane with tunneling resistance $R_{T}=26$ k$\mathrm{\Omega}$ and bottom curve for 30 nm thick membrane with $R_{T}=9$ k$\mathrm{\Omega}$. (b) Calculated gauge factors for the 65 nm (black squares) and the 30 nm data (open circles) in (a). (c) Relative change in tunneling resistance vs. calculated strain $|\epsilon_{zz}|$ for the 65 nm data in (b), (d) same for the 30 nm data. Red solid lines in (c) and (d) are calculations based on Eq. (\ref{response}) with $\tilde{\phi}_{0}$ and $d$ as free parameters.}
\end{figure}

Examples of the measured response of the tunneling resistance $R_T$ vs. displacement (a lateral distance 1.5 $\mu$m away from the location of the tunnel junction) for two different samples is shown in Fig. \ref{fig:response}(a). We have measured several more samples with essentially identical response in the normalized units $R_T/R_0$, where $R_0$ is the unstrained resistance. Short--circuited samples showed no observable response, thus the observed signal is due to tunneling, not due to the piezoresistive behavior of the Al metal film \cite{Piezo2}. The top curve is for a 65 nm thick membrane with $R_{T}=26$ k$\mathrm{\Omega}$, while the lower curve  is data for a 30 nm thick membrane  and $R_{T}=9$ k$\mathrm{\Omega}$. Black solid lines correspond to the pressing and open red circles to the release of the membrane.  
The origin of the reproducible hysteresis seen for the thin 30 nm membranes is mechanical, as it was also verified by the FZ curves (not shown here). Detailed understanding of the hysteresis is currently lacking.

To gain more quantitative understanding of the strain responsivity of a tunnel junction, we discuss here a model based on the widely used simple trapezoidal barrier picture (Simmons model) for a tunnel junction \cite{Simmons} and an isotropic elastic solid.    
The change in the tunneling resistance is caused by a deformation of the tunneling barrier, which manifests itself as a change of the effective barrier thickness $d$, the barrier height $\phi_{0}$ and the area $A$ of the junction. From geometrical arguments one can easily write how the strained values of $d$ and $A$ depend on the compressive (diagonal) strain components $\epsilon_{\{xx,yy,zz\}}$ in our geometry of a parallel plate junction in the x-y plane (plane of the membrane) with the tunneling current along the z-coordinate: $\tilde{d}=d(1+\epsilon_{zz})$, $\tilde{A}=A(1+\epsilon_{xx}+\epsilon_{yy})$. The shear strain components will not have an effect in the lowest order.  Thus, the strained tunneling resistance $\tilde{R}_{T}$ at zero bias voltage is 
%\begin{widetext}
\begin{equation}
\tilde{R}_{T}=\frac{h^2}{e^2\sqrt{2m\tilde{\phi_{0}}}}\frac{d}{A}
\frac{1+\epsilon_{zz}}{1+\epsilon_{xx}+\epsilon_{yy}}
\exp\left[\frac{2d}{\hbar}\sqrt{2m\tilde{\phi_{0}}}(1+\epsilon_{zz})\right]
\ ,
\label{response}
\end{equation}  
%\end{widetext}

%\begin{widetext}
%\begin{equation}
%\tilde{R}=\frac{h^2}{e^2\sqrt{2m\tilde{\phi_{0}}}}\frac{1+\epsilon_{z}-\nu(\epsilon_{x}+\epsilon_{y})}
%{1+\epsilon_{x}+\epsilon_{y}-\nu(\epsilon_{x}+\epsilon_{y}+2\epsilon_{z})}
%\exp\left\{\frac{2d}{\hbar}\sqrt{2m\tilde{\phi_{0}}}[1+\epsilon_{z}-\nu(\epsilon_{x}+\epsilon_{y})]\right\} \ ,
%\end{equation}  
%\end{widetext}
 where $\tilde{\phi_{0}}$ is the average barrier height in the strained configuration, changing because of the variation of the band gap with strain (deformation potential theory) \cite{bardeen}.   At first sight the dependence on the plain strains $\epsilon_{\{xx,yy\}}$ seems weak; However, from the elasticity theory for isotropic solids we can derive that $\epsilon_{zz}=T_{zz}(1-\nu-2\nu^2)/[(1-\nu)E]-(\epsilon_{xx}+\epsilon_{yy})\nu/(1-\nu)$,  where $\nu$ is the Poisson ratio, $E$ the Young's modulus and $T_{zz}$ the stress perpendicular to the membrane. This means that even in the normal--stress--free regions $T_{zz}=0$ of the membrane (far from the tip position) $\tilde{R}_{T}$ still has responsivity due to the in-plane strains (effect of the finite Poisson ratio).  

One important figure--of--merit in strain sensors is the gauge factor $\gamma$, which is defined as a ratio of the relative change in the resistance to the applied strain $\epsilon$, i.e. $\gamma=(\Delta R/R_{0})/\epsilon$. In our experiment, the direct measurement of the local strain at the tunnel junction location is impossible, as no nanoscale commercial strain gauges exist.    % In our setup, the direct measurement of the strain with an additional, calibrated strain gauge was impossible due to the small size of the system. 
However, we have calculated the strain fields in our experimental geometry (SiN membrane and Al wire) using the finite element method (FEM) with the displacement at the tip location as the boundary condition \cite{note2}.  All simulations were performed using non--linear strain (large deflection theory), as the displacements in the experiments are much larger than the membrane thickness. 

As we can see from Eq. (\ref{response}), the tunnel junction responds mainly through $\epsilon_{zz}$, so that in our case the most reasonable way to define a gauge factor is with that strain component.  
The calculated gauge factors for a number of membrane displacements are shown in Fig. \ref{fig:response}(b), using the experimentally determined resistance response from the data in Fig. \ref{fig:response}(a). 
The 65 nm membrane shows a peak value of $\gamma=37$ at a displacement of 1.5 $\mu$m, and a slightly declining trend with increasing displacement, whereas the 30 nm values are clearly lower with a maximum of $\gamma=19$, but with a different, decreasing trend with the displacement. The values of $\gamma$ are about an order of magnitude higher than for typical thin metal films such as gold \cite{Piezo2,Roukes}. The overall difference between the $\gamma$ values of the two membranes can be explained by the difference in $R_T$, which results from differences in $d$ and $\phi_0$.  

With the simulated strain, we can also study how the measured tunneling resistance varies with $\epsilon_{zz}$, instead of just displacement. 
Figures \ref{fig:response} (c) (65 nm membrane) and (d) (30 nm membrane) show  this for the same strain simulations as were used in in Fig. \ref{fig:response}(b). We see that the response is fairly linear in both cases, and that the value of strain is much larger for the thinner membrane. The quasi--linearity of the response may seem surprising, based on Eq. (\ref{response}). We point out, however, that as the absolute values of strain are $\sim 10^{-4}$, the exponent is small enough to behave linearly. This is verified by plotting theoretical points based on the computed strain and the modelled resistance response from Eq. (\ref{response}) in Figs. \ref{fig:response} (c) and (d), also.   
% even though the response to displacement is quite non--linear (Fig. \ref{fig:response}(a)). This happend because of the strong non--linearity of the strain--to--displacement relation in the large deflection limit \cite{Timoshenko}.
Moreover, the different dependence of the gauge factor on the displacement can also be explained by our model, Eq. (\ref{response}). %This is clear from the theoretical points in Fig. \ref{fig:response}(b). 
In all the theoretical response calculations, we estimated the effect of $\tilde{\phi}_{0}$ by using a typical value for the deformation potential constant in semiconductors $a=-4$ eV, where the change of energy gap $\Delta E_g=a(\epsilon_{xx}+\epsilon_{yy}+\epsilon_{zz})$ \cite{bardeen}. The accurate value for AlO$_{\mathrm{x}}$ is not known to us.     

\begin{figure}
\includegraphics[width=8.5cm]{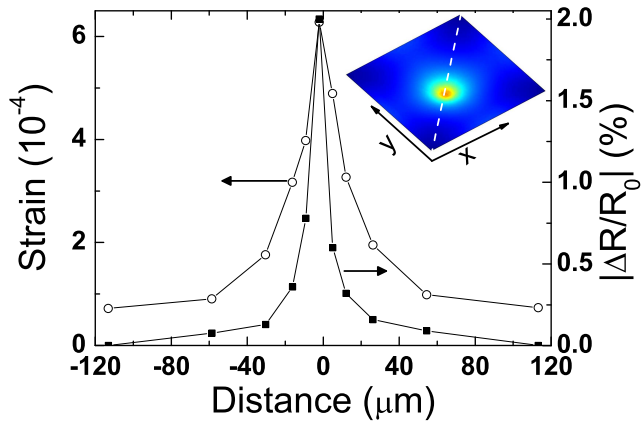}
\caption{\label{fig:spatial} (Color online) Calculated strain $|\epsilon_{zz}|$ (open circles) and measured relative change in the tunneling resistance (black squares) as a function of distance from the center of the membrane through the diagonal of the membrane (white dashed line in the inset). Inset shows the calculated strain field, when membrane is actuated in the middle.}
\end{figure}

In addition to local measurements, we have investigated the spatial dependence of the response by actuating the membrane at different points in the plane ($x$--$y$ coordinates) of the membrane, and measuring the response at a fixed location. The results for the experimental response and for calculated $\epsilon_{zz}$ are shown in Fig. \ref{fig:spatial}(a), where the actuation point was moved along the diagonal  of the membrane (dashed white line in Fig. \ref{fig:spatial} inset).  The inset shows the shape of the simulated strain field when the center of the membrane is pressed. Correspondence between the strain field and the tunnel junction response is good at a qualitative level, although some differences remain that we do not yet fully understand.

In summary, we have shown that  sub--micron sized Al-AlO$_{\textrm x}$-Al tunnel junctions can be used for strain sensing in low--dimensional nanostructures, yielding relatively high gauge factors. By measuring the resistance more accurately with a realistic sensitivity $\Delta R/R \sim 10^{-6}$, static strains down to $10^{-7}$ could be measured with $\gamma\sim10$. That would correspond to a static displacement sensitivity of 0.4 pm for a cantilever of thickness 1 $\mu$m and length 1 $\mu$m at the cantilever end. We also showed that a simple model of the tunnel barrier can explain the response behavior well, and can be used to predict that a higher barrier would increase the responsivity. For typical AlO$_{\textrm x}$ tunnel barriers, $\phi_0 \sim$ 1 eV \cite{anniilaa}, which is well below the bulk value closer to 10 eV. Thus if one could find a way to make barriers closer to the bulk values, the responsivity would increase by a factor three. Low temperature measurements would also help in reducing the noise generated by the junction,   and would provide more information on the microscopic behavior of AlO$_{\mathrm{x}}$ barriers under strain, because information of the charging energy (capacitance)  variation could then be measured \cite{anniilaa}.
We thank Thomas K\"uhn for useful discussions. This work has been supported by the Academy of Finland under projects 128532 and 118231. J.T.L. acknowledges the National Graduate School in Materials Physics for financial support.
%P.J.K. acknowledges the National Graduate School in Materials Physics for partial financial support.
%\bibliography{AFMmembrane}% Produces the bibliography via BibTeX.

\begin{thebibliography}{24}
\bibitem[{not({\natexlab{a}})}]{note}
\bibinfo{journal}{In our experiments we used two different types of
  cantilevers: for thinner membrane (thickness 30 nm) OMCL--AC160TS probes from
  Olympus with a nominal spring constant of 42 N/m and for thicker (thickness
  65 nm) MPP--13100 probes from Digital Instruments Veeco with a spring
  constant of 200 N/m.}
\bibitem[{not({\natexlab{b}})}]{note2}
\bibinfo{journal}{The thin AlO$_{\mathrm{x}}$ barrier could not be included in
  the computation due to computer memory limitations. However, we estimate that
  its contribution is small. Material parameters used in the simulations were
  taken from Ref. \cite{ClelandNEMS}, with Young's modulus $E=357$ GPa and
  Poisson ratio $\nu=0.25$ for SiN, and $E=70$ GPa and $\nu=0.35$ for amorphous
  Al. Simulations were performed using commercial software Comsol Multiphysics
  from Comsol AB.}
\end{thebibliography}

\end{document}